\documentclass[a4paper,aps,prd,10pt,preprintnumbers,showpacs,twocolumn,superscriptaddress,nofootinbib,amsmath,amssymb]{revtex4-1}
\usepackage{graphicx}
\usepackage[utf8]{inputenc}
\usepackage[T1]{fontenc}
\usepackage{cmap}
\usepackage{ulem}

\begin{document}

\title{Exploring Unique Quasinormal Modes of a Massive Scalar Field in Brane-World Scenarios}
\author{Antonina F. Zinhailo}
\email{F170631@fpf.slu.cz}
\affiliation{Institute of Physics and Research Centre of Theoretical Physics and Astrophysics, Faculty of Philosophy and Science, Silesian University in Opava, CZ-746 01 Opava, Czech Republic}

\begin{abstract}
We compute precise values of quasinormal modes of a massive scalar field in the background of the Schwarzschild-like brane-localised black holes. It is shown that the quasinormal spectrum of the massive field differs qualitatively from that previously known for other black hole models, due to the presence of two kinds of modes: those whose damping rate vanishes as the mass of the field $\mu$ increases up to some critical value, and those whose real oscillation frequency vanishes at a certain value of $\mu$. While the first type of modes, which are arbitrarily long-lived, are recognized in various four-dimensional backgrounds as quasi-resonances, the second type is a novel feature for asymptotically flat black holes. When $Re (\omega)$ reaches zero, the fundamental mode disappears from the spectrum and the first overtone becomes the fundamental mode. We also demonstrate that quasi-resonances may not exist for brane-localised black holes immersed in $D\geq 6$ - dimensional bulk.   
\end{abstract}


\maketitle

\section{Introduction}

A hierarchy problem arises from attempts to understand why the characteristic scale of gravity, $M_{P} \sim 10^{19} GeV$, is orders of magnitude larger than the Electro-Weak scale ($\sim TeV$). An elegant solution to this problem was proposed by considering the existence of large extra dimensions, where the graviton can propagate throughout the entire $D$-dimensional spacetime, while matter fields are confined to our $(3+1)$-dimensional brane \cite{Arkani-Hamed:1998jmv,Antoniadis:1998ig,Randall:1999ee,Randall:1999vf}.

In such models, and to a large extent independently of the particular brane-world scenario chosen, black holes observed from our brane can be effectively described in a universal manner, provided that their event horizon radius $r_0$ is significantly smaller than the size of the large extra dimension $L$. The black hole metric is governed by the exact (Tangherlini) solution of the $D$-dimensional Einstein equations, where the $D-2$ spherical element is replaced by the usual two-dimensional sphere. The upper limit on the size of extra dimensions is determined by current tests of Newton's law, which are accurate at distances approximately above 1 mm.   

Quantum (Hawking) and classical (driven by the proper oscillation frequencies, quasinormal modes, \cite{Konoplya:2011qq,Kokkotas:1999bd}) radiation around brane-localised black holes have been extensively studied in \cite{Kanti:2005xa,Kanti:2006ua,Berti:2003yr,Zhidenko:2008fp,Chung:2015mna,Harris:2003eg,Kanti:2004nr}. However, while Hawking radiation and grey-body factors for a massive scalar field around brane-localised black holes were investigated in \cite{Kanti:2010mk}, there has been no such study on quasinormal modes, which was limited by massless fields in this case, to the best of our knowledge. However, recently the asymptotic tails of the massive scalar field in the brane-localised black holes \cite{Dubinsky} were studied, where it is was shown that the asymptotic fall-off at $t \rightarrow \infty$ differs from that for the Schwarzschild case.

Simultaneously, the study of massive fields has its own motivation. Effective massive terms emerge when perturbing a massless field in various models of higher-dimensional gravity \cite{Seahra:2004fg,Ishihara:2008re}, or when a black hole is immersed in a magnetic field \cite{Konoplya:2008hj,Konoplya:2007yy}. The late-time decay of massive fields is characterized by oscillatory tails, extensively studied in \cite{Jing:2004zb, Koyama:2001qw, Moderski:2001tk, Konoplya:2006gq, Rogatko:2007zz, Koyama:2000hj, Churilova:2019qph, Gibbons:2008rs, Gibbons:2008gg,Konoplya:2024ptj}, with potential observations discussed within the Timing Pulsar Array experiments \cite{Konoplya:2023fmh}.

An interesting property of the quasinormal spectrum of massive fields is the appearance of arbitrarily long-lived modes, termed {\it quasi-resonances} \cite{Ohashi:2004wr,Konoplya:2017tvu}. The damping rate of such modes, determined by the imaginary part of the complex quasinormal frequency $\omega$, slowly decreases until reaching zero. At this point, the mode, previously fundamental, disappears from the spectrum, while the first overtone becomes the new fundamental mode. This phenomenon of arbitrarily long-lived modes has been investigated for various black hole backgrounds and spin configurations of fields \cite{Konoplya:2019hml,Bolokhov:2023bwm,Bolokhov:2023ruj,Zinhailo:2018ska,Konoplya:2005hr,Konoplya:2017tvu}.

In all the aforementioned examples of asymptotically flat black holes, evidence of the existence of quasi-resonances has been demonstrated, with a sole counter-example being the Schwarzschild-de Sitter black hole \cite{Konoplya:2004wg}, where an analytical proof was provided showing that quasi-resonances are not permitted due to the non-zero cosmological constant. Therefore, the presence of quasi-resonances in the spectrum of a massive field around an asymptotically flat black hole was anticipated, though not rigorously proven in the general case. In this paper, we will present numerical indications suggesting that quasi-resonances may not exist for brane-localised black holes immersed in $D \geq 6$ bulk. 

With the aforementioned motivations in mind, we will study the quasinormal modes of a massive scalar field, considering it as the simplest qualitative model also for higher spin particles. This supposition is further supported by the observation that the spectrum in the high-frequency (eikonal) regime does not depend on the spin of the field, although this is not always the case \cite{Konoplya:2017wot,Bolokhov:2023dxq,Konoplya:2022gjp,Konoplya:2019hml}.

Here, we will demonstrate that the spectrum on the brane is essentially determined by the dimensionless quantity $\mu M/m_{P}^2$, where $\mu$ is the mass of the particle, $M$ is the black hole mass, and $m_{P}$ is the four-dimensional Planck mass. Considering that for Standard Model particles $\mu \ll m_{P}$, our analysis would be reliable not only for very light black holes on the order of the Planck mass, but also for relatively large masses of black holes, which still fall within the range $\mu M/m_{P}^2 \lesssim 1$. Here, we will utilize the convergent Frobenius method for the analysis of quasinormal modes, without implying any constraints or approximations for the mass of the field or the black hole.

We will demonstrate that the spectrum of a massive scalar field propagating in the vicinity of the brane-localised black hole differs qualitatively from the four-dimensional black holes studied thus far, as well as from the scalar field propagating in the entire $D$-dimensional spacetime.

Our letter is organized as follows: In sec. II, we summarize the basic information on the black hole metric and the wave-like equation. Sec. III briefly reviews the methods used for calculations of the quasinormal frequencies and discusses the obtained numerical data. In the conclusion, we summarize the findings and discuss some open questions.

\section{The black hole metric and the wave equation}

\begin{figure*}
\resizebox{\linewidth}{!}{\includegraphics{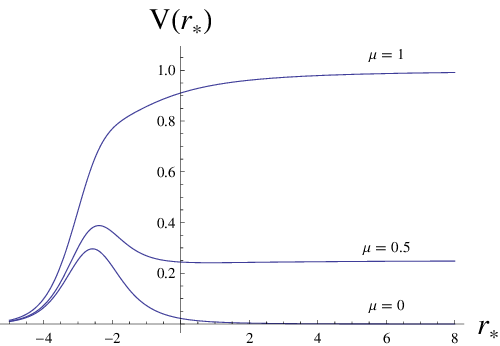}\includegraphics{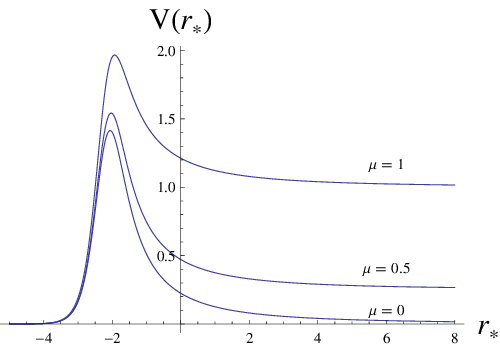}}
\caption{Effective potentials for $\ell=0$, $D=5$ (left) and $\ell=1$, $D=7$ (right) scalar field perturbations for various values of $\mu$.}\label{fig0}
\end{figure*}

The metric of a static spherically-symmetric brane-localized  black hole is given by the following line element,
\begin{equation}\label{metric}
ds^2=-f(r)dt^2+\frac{dr^2}{f(r)}+r^2 (d\theta^2+\sin^2\theta d\phi^2),
\end{equation}
where  the metric function of the $D$-dimensional black hole projected onto the $3+1$-dimensional brane \cite{Kanti:2004nr,Kanti:2006ua} is
\begin{equation}
f(r) = 1- \left(\frac{r_{0}}{r}\right)^{D-3} = 1- \frac{2 M}{r^{D-3}}.
\end{equation}
Here $M$ is the mass parameter. The massive matter fields are supposed to be propagating within the $(3+1)$-dimensional brane, while the gravitational field is propagting into the bulk as well.
The massive Klein-Gordon equation in a curved spacetime
\begin{equation}
\frac{1}{\sqrt{-g}}\partial_\mu \left(\sqrt{-g}g^{\mu \nu}\partial_\nu\Phi\right) - \mu^2 \Phi =0,
 \end{equation}
can be reduced to the following master wave-like equation
\begin{equation}\label{wave-equation}
\dfrac{\partial^2 \Psi}{\partial r_*^2} - \frac{\partial^2 \Psi}{\partial t^2} -V(r) \Psi=0,
\end{equation}
once the separation of variables is performed and the new wave function is intoroduced. Here the ``tortoise coordinate'' $r_*$ is defined as follows:
\begin{equation}
dr_*\equiv\frac{dr}{f(r)}.
\end{equation}
The effective potential has the  form:
\begin{equation}
V(r) = f(r) \left(\frac{\ell(\ell+1)}{r^2}+\frac{1}{r}\frac{d f(r)}{dr} +  \mu^2\right),
\end{equation}
where $\ell$ is the multiipole number. Here we will use the units of the event horizon radius $r_{0}$, implying that $\mu$ is in fact $\mu r_{0}$ and $\omega$ is $\omega/r_{0}$ in dimensionless units. The examples of the effective potentials as a function of the tortoise coordinate are shown in figs. \ref{fig0}.

\section{Quasinormal frequencies}

\begin{figure*}
\resizebox{\linewidth}{!}{\includegraphics{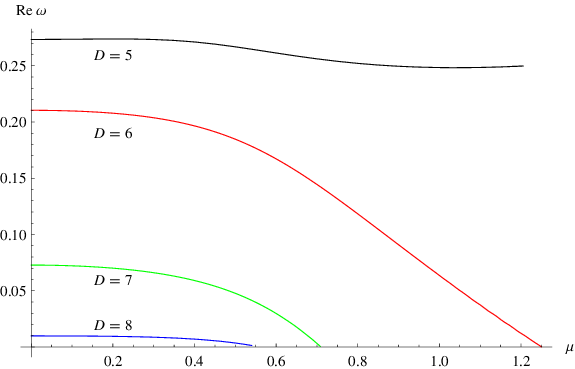}\includegraphics{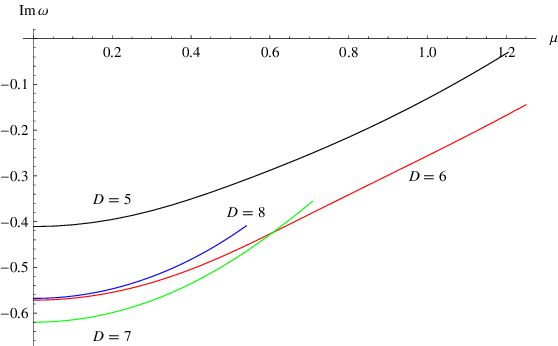}}
\caption{Real (left) and imaginary (right) parts of the fundamental quasinormal mode ($n=0$), $\ell=0$, $r_{0}=1$ for various $D$ as a function of the field's mass $\mu$. }\label{fig1}
\end{figure*}

\begin{figure*}
\resizebox{\linewidth}{!}{\includegraphics{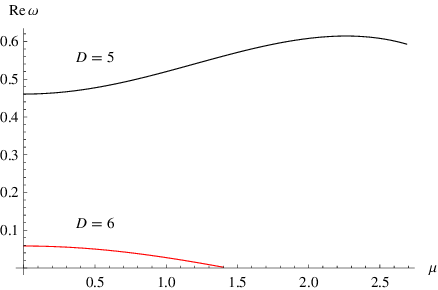}\includegraphics{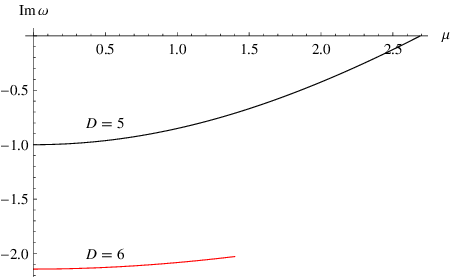}}
\caption{Real (left) and imaginary (right) parts of the first overtone  ($n=1$), $\ell=0$, $r_{0}=1$ for $D=5$ and $6$ as a function of the field's mass $\mu$.}\label{fig2}
\end{figure*}

\begin{figure*}
\resizebox{\linewidth}{!}{\includegraphics{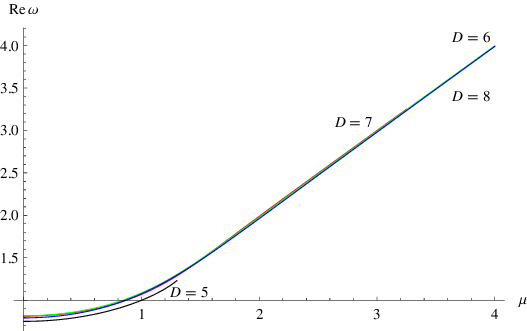}\includegraphics{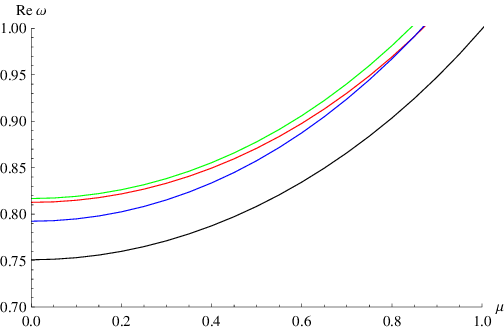}\includegraphics{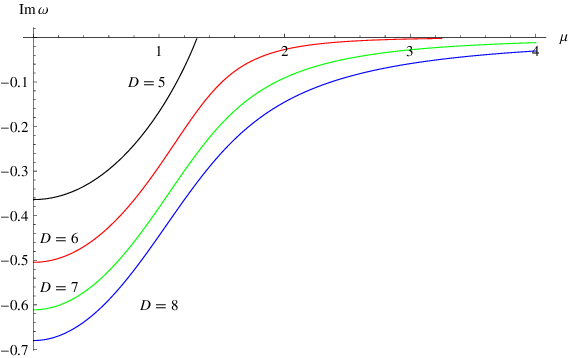}}
\caption{Real (left) and imaginary (right) parts of the fundamental mode  ($n=0$), $\ell=1$, $r_{0}=1$ for various $D$ as a function of the field's mass $\mu$. The left plot is enlarged in the middle figure, where $D =5, 6, 7, 8$ from bottom to top.}\label{fig3}
\end{figure*}

\begin{figure*}
\resizebox{\linewidth}{!}{\includegraphics{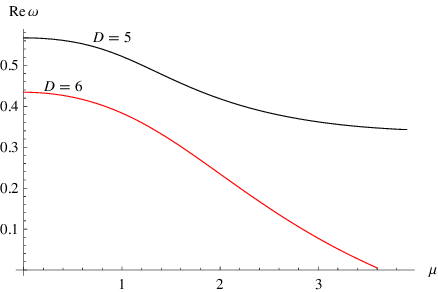}\includegraphics{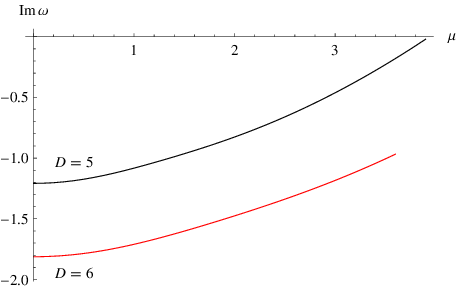}}
\caption{Real (left) and imaginary (right) parts of the first overtone  ($n=1$), $\ell=1$, $r_{0}=1$ for $D=5$ and $6$ as a function of the field's mass $\mu$.}\label{fig4}
\end{figure*}

We will use the Frobenius or Leaver method for the analysis of frequencies of a massive scalar field. This method is based on the convergent series expansion allowing one to determine the quasinormal modes with any desired precision ~\citep{Leaver:1985ax,Leaver:1986gd}. For stronger convergence we also use the Nollert improvement \citep{Nollert:1993zz}, which was generalized in \citep{Zhidenko:2006rs} to an arbitrary number of terms of the recurrence  relations.

The wave-like equation has always a regular singular point at the event horizon $r=r_0$ and the irregular singular point at $r=\infty$.
The new radial function $P(r, \omega)$, is introduced in such a way 
\begin{equation}\label{reg}
\Psi(r)= P (r, \omega) y(r),
\end{equation}
that the factor $P(r, \omega)$  provides regularity of $y(r)$ in the range $r_0\leq r$ at the quasinormal modes boundary conditions.
Then, $y(r)$ can be represented in the form of a following series:
\begin{equation}\label{Frobenius}
y(r)=\sum_{k=0}^{\infty}a_k\left(1-\frac{r_0}{r}\right)^k.
\end{equation}
We use the Gaussian eliminations or a system of linear equations and, further, reduce finding of $\omega$ via numerical solution of a non-algebraic equation with the help of the FindRoot command in {\it Mathematica}.
For the latter we need an initial guess for the root, and for this purpose we used the 6th order WKB formula \cite{Konoplya:2003ii} with Pade approximants \cite{Matyjasek:2017psv}. The automatic WKB code we used was shared in \cite{Konoplya:2019hlu}. While the WKB method works relatively well for guessing the fundamental mode at lower $D=4,5,6$, at higher $D$ it is unstable in the sense that a change of the WKB order or Pade split by one change the results significantly. In this case we use the time-domain integration method in the form suggested in \cite{Gundlach:1993tp} used in numerous subsequent works (see, for example, \cite{Konoplya:2014lha,Bolokhov:2023dxq,Varghese:2011ku}).  It is essential that when the singular points  appear within the unit circle $|x| < 1$ (where $x$ is a compact coordinate for which $x=0$ corresponds to the event horizon and $x=1$ to inifnity), we employ integration though a sequence of positive real midpoints using the  Rostworowski approach \cite{Rostworowski:2006bp}.

\begin{table*}
\begin{tabular}{|c|c|c|c|}
  \hline
  \hline
$\mu$  &  $\omega$ ($\ell=0$) & $\omega$ ($\ell=1$) & $\omega$ ($\ell=2$)  \\
  \hline
0   & 0.273387 - 0.410907 i & 0.750847 - 0.363873 i & 1.250161 - 0.357430 i\\
0.2 & 0.273809 - 0.395262 i & 0.759893 - 0.356502 i & 1.257118 - 0.354331 i \\
0.5 & 0.266843 - 0.320650 i & 0.808348 - 0.317237 i & 1.293802 - 0.338108 i \\
0.7 & 0.256140 - 0.253055 i & 0.865980 - 0.271061 i & 1.336060 - 0.319667 i \\
1 &   0.248386 - 0.130999 i & 0.999379 - 0.167038 i & 1.427097 - 0.280879 i\\
  \hline
  \hline
\end{tabular}
\caption{The fundamental ($n=0$) quasinormal modes for $D=5$ and various $\mu$.}
\end{table*}

\begin{table*}
\begin{tabular}{|c|c|c|c|}
  \hline
  \hline
$\mu$  &  $\omega$ ($\ell=0$) & $\omega$ ($\ell=1$) & $\omega$ ($\ell=2$)  \\
  \hline
0   & 0.210401 - 0.571674 i & 0.812651 - 0.504544 i & 1.399940 - 0.498226 i\\
0.2 & 0.207824 - 0.554624 i & 0.821745 - 0.495764 i & 1.407161 - 0.494518 i \\
0.5 & 0.184702 - 0.467971 i & 0.870897 - 0.449675 i & 1.445213 - 0.475260 i \\
0.7 & 0.144494 - 0.384196 i & 0.930243 - 0.397263 i & 1.488979 - 0.453698 i \\
1 &  0.063583 - 0.256020 i & 1.069406 - 0.289511 i & 1.582903 - 0.409568 i\\
  \hline
  \hline
\end{tabular}
\caption{The fundamental ($n=0$) quasinormal modes for $D=6$ and various $\mu$.}
\end{table*}

First, we will discuss the $\ell=0$ case, which exhibits qualitatively distinctive features from higher $\ell$ values. In Fig. \ref{fig1}, we observe that the behavior of the fundamental ($n=0$) mode differs qualitatively for $D=5$ compared to $D=6$ and higher numbers of spacetime dimensions. The $D=5$ case resembles the four-dimensional scenario \cite{Ohashi:2004wr,Konoplya:2004wg}: the damping rate of the fundamental mode, represented by $Im (\omega)$, tends to zero as $\mu$ increases, while the real oscillation frequency approaches some constant. When the damping rate approaches zero, the fundamental mode vanishes from the spectrum, and the first overtone becomes the new least damped mode.

In contrast, $D=6$ and higher cases exhibit peculiar behavior: the oscillation frequency $Re (\omega)$ approaches zero as $\mu$ increases, and although the damping rate also decreases, there is no evidence that it reaches zero. This is due to the slow convergence of the Leaver method, especially when frequencies have tiny real parts. There is indication that the mode disappears from the spectrum when $Re(\omega)$ reaches zero at some non-zero $Im (\omega)$. Indeed, considering that for $D=7$ and $D=8$, the damping rate remains quite large for almost vanishing values of $Re (\omega)$, it is probable that the damping rate does not vanish at the critical value of $\mu$ for which $Re (\omega) \rightarrow 0$.

As depicted in Fig. \ref{fig1}, the fundamental mode vanishes from the spectrum, and the first overtone, illustrated in Fig. \ref{fig2}, becomes the new fundamental mode. It is noteworthy that when the mass of the field is zero, our results align with those published in Table I of \cite{Kanti:2006ua}. For easy verification of the obtained quasinormal modes, we also provide Tables I and II, which includes frequencies for $\ell = 0, 1$, and $2$ for a few values of $\mu$.

It is noteworthy that modes with zero real part are known even for massless fields. Specifically, in gravitational perturbations of the Schwarzschild solution, there exist purely imaginary quasinormal modes referred to as algebraically special \cite{Couch:1973zc}, which cannot accurately represent the real perturbation process. However, in our $D \geq 6$ case, $Re (\omega)$ approaches zero only asymptotically. Thus, the modes under consideration evidently do not fall within the class of algebraically special ones.

The quasinormal spectrum of the massive scalar field for $\ell =0$ exhibits qualitative differences not only from the well-known four-dimensional cases but also from $D>5$ black holes that are not projected onto the brane, whose spectrum was examined in \cite{Zhidenko:2006rs}. The $\ell=0$ modes for $D$, as presented in Fig. 1 of \cite{Zhidenko:2006rs}, have a real part approaching the field's mass, whereas in our case, it approaches zero::
\begin{align}\nonumber
Re (\omega) \rightarrow \mu, \quad D \geq 5, \quad \ell=0, \quad bulk\\ 
Re (\omega) \rightarrow 0, \quad D>5, \quad \ell=0, \quad brane.
\end{align}

The case of $\ell=1$ and higher is also different from the four dimensional case, but similar to the higher dimensional case of non-projected black holes. In this case for the fundamental mode $n=0$ we have (see fig. \ref{fig3})  
\begin{align}\nonumber
Re (\omega) \rightarrow const \neq 0, \quad Im (\omega) \rightarrow 0, \quad \mu \rightarrow \mu_{crit} \quad (D =5),\\ \nonumber
Re (\omega) \rightarrow \mu, \quad Im (\omega) \rightarrow 0, \quad \mu \rightarrow \infty \quad (D \geq 6).
\end{align}

The first overtones $n=1$ for $\ell=1$ and $D=5$ and $6$ are shown in fig. \ref{fig4}. There we see that the behavior is similar to the fundamental mode, that is, for $D=5$ the first overtone becomes quasi-resonance at some $\mu$, while for $D=6$ the real oscillation frequency vanishes at some $\mu$ and the overtone does not reach zero damping rate. Thus, we conclude that for black holes projected on the brane arbitrarily long lived modes are allowed only for $D=5$.

Unlike the case of propagation in the $D$-dimensional spacetime considered in   \cite{Zhidenko:2006rs}, for black holes projected on the brane, there is no evidence of the existence of quasi-resonances for $D \geq 6$. In the non-projected $D$-dimensional black holes, the first and higher overtones become quasi-resonances as was shown in   \cite{Zhidenko:2006rs}. In our case, on the contrary, the overtones behave similarly to the fundamental mode.

\section{Discussions}
From \cite{Argyres:1998qn} one can see that the radius of the event horizon $r_{0}$ and the mass of the black hole are related as follows
\begin{equation}
r_{0} \sim \frac{1}{M_{*}} \left(\frac{M}{M_{*}}\right)^{\frac{1}{D-3}}, 
\end{equation}
where the factor due to the solution of the D-dimensional Einstein equation is not written out explicitly here. The quantity $M_{*}$ determines the fundamental Planck scale, which is different from the four-dimensional one $m_{P}$: 
\begin{equation}
m_{P}^2 = L^{D-4} M_{*}^{D-2}.
\end{equation}
For $D=5$, for example, this implies that the fundamental Planck scale could be a few orders or more larger than the 4-dimensional one. 

Implying, further, that our black holes,  could not be bigger than $r_{0} \sim 10^{-5}$ m., which is two orders smaller than the maximal size of the extra dimension,  we find that the dimensionless product of the particle mass and mass of such black hole $\mu M/m_{P}^2$ could be around unity, where all our interesting effects come into play, for $M \sim 10^{15}-10^{16}$  kg. At $\mu M/m_{P}^2 \sim 1$, this roughly corresponds to $\mu \sim 10^{-30}$ kg., which is of the order of the electron mass and some other particles.

In the present paper we have shown that the spectrum of  massive scalar field in the background of the simplest Tangherlini black hole projected on the brane is qualitatively different from those for other known four-dimensional black holes as well as from higher dimensional black holes where the scalar field propagates in all $D$-spacetime dimensions. These distinctions are:
\vspace{2mm}

\begin{itemize}
\item The real part of the frequency  tends to zero at some critical value of the mass $\mu$ for (a) the fundamental mode of $\ell=0$  at $D \geq 6$,  and (b)  overtones of $\ell=1, 2,..$   at $D \geq 6$, 
\item The real part of the frequency for the fundamental mode of  $\ell=1, 2,..$ at $D \geq 6$ tends to $\mu$ when $\mu \rightarrow \infty$,
\item There is no evidence of the existence of quasi-resonances for $D \geq 6$.
\end{itemize}

It would be intriguing to extend our work to the case of a non-zero cosmological constant. General spherically symmetric and asymptotically de Sitter black holes do not permit arbitrarily long-lived modes of a massive scalar field, as analytically demonstrated in \cite{Konoplya:2004wg}. Consequently, the first type of modes must be absent. The intriguing question arises: what would happen with the second type of modes existing at $D \geq 6$, where the real oscillation frequency vanishes, once the cosmological constant is introduced?

Another characteristic of the spectrum of asymptotically de Sitter black holes is the presence of two branches of modes: one derived from the Schwarzschild solution deformed by the introduction of the cosmological constant \cite{Konoplya:2004uk,Zhidenko:2003wq}, and the other stemming from the empty de Sitter space distorted by the presence of the black hole \cite{Lopez-Ortega:2009jpx,Cardoso:2017soq}. The second branch consists of purely imaginary, i.e., non-oscillatory frequencies \cite{Konoplya:2022xid}. This, coupled with the effects related to the mass of the field considered in the present work,  promises a rich and complex structure of the quasinormal spectrum.

Among our future plans is to generalize our considerations to the case of non-zero charge and angular momentum.

\acknowledgments
I would like to thank R. A. Konoplya and A. Zhidenko for useful discussions and help.
I also acknowledge the Silesian University grant SGS 2024 for support. 

\bibliographystyle{unsrt}
\bibliography{Bibliography}

\end{document}